\documentclass[pra,twocolumn,showpacs,floatfix]{revtex4}

\usepackage{amsmath}
\usepackage{times}
\usepackage{graphicx}
\usepackage{color}

\begin{document}

\title{Line shapes of optical Feshbach resonances near the intercombination transition of bosonic Ytterbium}
\author{M. Borkowski$^{1}$, R. Ciury\l{}o$^{1}$, P. S. Julienne$^{2}$,
S. Tojo$^{3}$, K. Enomoto$^{4}$, Y. Takahashi$^{5}$}
\affiliation{$^{1}$ Instytut Fizyki, Uniwersytet Miko\l{}aja
Kopernika, ul. Grudzi\c{a}dzka 5/7, 87--100 Toru\'n, Poland.\\
$^{2}$ Joint Quantum Institute, National Institute of Standards and
Technology  and The University of Maryland,
100 Bureau Drive, Stop 8423, Gaithersburg, Maryland 20899-8423, USA.\\
$^{3}$Department of Physics, Faculty of Science, Gakushuin University,
Tokyo 171-8588, Japan.\\
$^{4}$Department of Physics, University of Toyama, 3190 Gofuku, Toyama 930-8555, Japan.\\
$^{5}$Department of Physics, Graduate School of Science, Kyoto
University, Kyoto 606-8502, Japan.}
\date{\today}

\begin{abstract}
The properties of bosonic Ytterbium photoassociation spectra near the intercombination transition
$^{1}S_{0}$--$^{3}P_{1}$ are studied theoretically at ultra low temperatures. We demonstrate how
the shapes and intensities of rotational components of optical Feshbach resonances are affected
by mass tuning of the scattering properties of the two colliding ground state atoms.
Particular attention is given to the relationship between the magnitude of the scattering length and
the occurrence of shape resonances in higher partial waves of the van der Waals system.
We develop a mass scaled model of the excited state potential that represents the experimental data
for different isotopes. The shape of the rotational photoassociation spectrum for various bosonic Yb
isotopes can be qualitatively different.
\end{abstract}

\pacs{34.50.Rk, 34.10.+x, 34.20.Cf, 32.80.Pj}

\maketitle

\section{Introduction}

The properties of intercombination transitions in alkaline earth atoms and atoms with similar
electronic structure has become an object of a growing number of experimental and theoretical studies.
It is mostly caused by a variety of new applications in the physics of ultra cold atoms: from
laser cooling and trapping \cite{ido03,curtis03,loftus04} to optical frequency standards
\cite{wilpers02,barber06,letargat06,boyd07}.

Great progress in this area has been achieved for Ytterbium (Yb), which has 7 stable isotopes with atomic
weights 168, 170, 171, 172, 173, 174, and 176.
Quantum degenerate gases have been obtained for the bosonic isotopes
$^{174}\rm{Yb}$ \cite{takasu03},  $^{170}\rm{Yb}$ \cite{fukuhara07a}, and $^{176}\rm{Yb}$ \cite{fukuhara09},
and the fermionic isotopes
$^{171}\rm{Yb}$ and $^{173}\rm{Yb}$\cite{fukuhara07b}. Photoassociation spectroscopy has been carried out
for bosons \cite{takasuPA, tojo06} as well as fermions \cite{enomoto08}.
A two-color photoassociation experiment
has allowed a precise determination of the ground state s-wave scattering lengths for all combinations
of Yb isotopes \cite{kitagawa08}. Finally, it was demonstrated that the significant change of the
scattering properties for colliding ground state atoms can be achieved with optical Feshbach
resonances in these systems \cite{enomoto08ofr}.

This work is devoted to the theoretical study of the photoassociation spectra near the intercombination
transition $^{1}S_{0}$--$^{3}P_{1}$ of Yb for bosonic isotopes. We take advantage
of the experimental photoassociation spectra for $^{174}\rm{Yb}_2$ and $^{176}\rm{Yb_2}$
from Tojo {\it et al}. \cite{tojo06} to precisely determine the binding energies of excited state molecular
energy levels.  In addition, we report new measurements of the binding energies of excited $^{170}\rm{Yb}_2$
and $^{172}\rm{Yb}_2$. Analysis of the photoassociation spectra also requires knowledge of the scattering
properties in the ground electronic state of two colliding atoms. The necessary information and experimental
data for all ground state isotopic combinations can be found in Ref. \cite{kitagawa08}.

Yb is an excellent example of a system for which the scattering properties can be easily tuned by the change
of the isotopic combination, thus changing the reduced mass of the colliding pair.
Such mass tuning of the scattering properties can also be very useful in other similar species with several
isotopes. Development in the laser trapping and cooling of atoms such as Cd \cite{brickman07}
and Hg \cite{walthe07,hachisu08} will hopefully allow photoassociation investigations of these
systems in the future.   Cadmium and Hg, like Yb, are good candidates for mass tuning of the
scattering length because of their numerous isotopes. In addition,
Hg is seen as a very promising candidate for future optical frequency standards.
The clock frequency shift induced by black body radiation is especially small in Hg
\cite{hachisu08}, compared with other Group II elements \cite{porsev06}.

\section{Photoassociation resonance}

The two-body loss rate coefficient $K(\Delta;I,T)$ in the photoassociation process for a thermal cloud
of ultracold atoms at temperature $T$ needs to be evaluated as an average over all possible momenta
of two colliding atoms. This loss rate is directly dependent on the light intensity  $I$ leading
to the photoassociation as well as on the detuning of the light from the atomic resonance $\Delta$
and detuning $\Delta_{e}$ corresponding to the optical resonance coupling the scattering ground state
"$g$" with the excited "$e$" bound state \cite{weiner99}.

The averaged loss rate can be written as \cite{ciurylo04}:
\begin{equation}
\label{equ1}
K(\Delta;I,T)=\left< {\cal K}(\Delta;I,\vec{p}_{c},\vec{p}_{r}) \right>_{\vec{p}_{c},\vec{p}_{r}}\,,
\end{equation}
where
\begin{widetext}
\begin{equation}
\label{equ2}
{\cal K}(\Delta;I,\vec{p}_{c},\vec{p}_{r})=
\frac{\hbar\pi}{k_{r}\mu}
\sum_{e,g}(2J_{g}+1) \frac{\Gamma_{pe}\Gamma_{eg}(I,\varepsilon_{r})}
{[\Delta + \varepsilon_{D}+\varepsilon_{r}-\Delta_{e}-E_{e}(I,\varepsilon_{r})-E_{\rm rec,mol}]^{2}+[\Gamma_{e}(I,\varepsilon_{r})/2]^{2}}
\end{equation}
\end{widetext}
is the loss rate \cite{napolitano94,julienne96,bohn99} corresponding
to particular momentum vectors of the relative motion of the two colliding
atoms $\vec{p}_{r}$ as well as the motion of their center of mass $\vec{p}_{c}$.
Contributions from all possible transitions between excited bound and ground scattering
states are included in this expression. They are taken in the sum with weights dependent
on the total angular momentum $J_{g}$ of the two-atom system.
In Eq. (\ref{equ2}) the magnitude of the wave vector corresponding
to the relative motion is $k_{r}=p_{r}/\hbar$,
the kinetic energy of relative motion is $\varepsilon_{r}={\hbar }^2 k_{r}^{2}/(2\mu)$,
and $\mu$ is the reduced mass of the colliding atoms. The Doppler shift is described by
$\varepsilon_{D}=-\hbar \vec{k}_{\rm las}\cdot\vec{p}_{c}/{\cal M}$ where
the magnitude of the laser light wave vector is $k_{\rm las}=\omega/c$ and the mass of the
molecule created in the photoassociation process is ${\cal M}$. The shift
of the photoassociation resonance $E_{\rm rec,mol}=\hbar^{2}k_{las}^{2}/(2{\cal M})$ caused by
the photon recoil is also included here. Finally, the light induced shift of a given
photoassociation resonance can be expressed as a sum
$E_{e}(I,\varepsilon_{r})=\sum_{g} E_{eg}(I,\varepsilon_{r})$ of contributions
$E_{eg}(I,\varepsilon_{r})$ of all possible optical transitions between the excited bound state
"$e$" and ground scattering states "$g$".  Similarly, the total width
of the resonance $\Gamma_{e}(I,\varepsilon_{r})=\Gamma_{pe}+\sum_{g}\Gamma_{eg}(I,\varepsilon_{r})$,
where $\Gamma_{eg}(I,\varepsilon_{r})$ is the light induced width between
the "$e$" and "$g$" states and $\Gamma_{pe}$ describes all possible other processes leading to loss of
the excited state. If radiative decay is the dominant loss process, then $\Gamma_{pe}$ can be taken as
the natural width of the excited molecular state.

The light induced width \cite{napolitano94,julienne96,bohn99}
\begin{equation}
\label{equ3}
\Gamma_{eg}(I,\varepsilon_{r})=2\pi\left|\left<\Psi_{e}\right|V_{\rm las}(I) \left|\Psi_{g}^{+}(\varepsilon_{r})\right>\right|^{2}
\end{equation}
is linearly dependent on the light intensity $I$ through the operator $V_{\rm las}(I)$ describing optical
coupling between particular excited and ground states. For the case investigated here more details about
this operator can be found in Refs. \cite{ciurylo04,napolitano97}. This width also depends on the kinetic
energy of the relative motion of the two colliding atoms $\varepsilon_{r}$. This is because the energy
normalized ground scattering state $\left|\Psi_{g}^{+}(\varepsilon_{r})\right>$ strongly depends
on this energy. Finally the magnitude of the light-induced width is dependent on the unit
normalized excited bound state $\left|\Psi_{e}\right>$.

The light induced shift \cite{bohn99} can be calculated from the Fano theory \cite{fano61,simoni02}
\begin{eqnarray}
E_{eg}(I,\varepsilon_{r})&=&\frac{1}{2\pi}{\cal P}\int_{0}^{\infty}d\varepsilon \frac{\Gamma_{eg}(I,\varepsilon_{r})}{\varepsilon_{r}-\varepsilon}
\nonumber\\
&&+\sum_{\left|\Psi_{g}\right>}\frac{\left|\left<\Psi_{e}\right|V_{\rm las}(I) \left|\Psi_{g}\right>\right|^{2}}{\varepsilon_{r}-E_{g}}\,,
\label{equ4}
\end{eqnarray}
where ${\cal P}$ is a principal part integral over all collision
energies, and the sum occurring in this expression is taken over all unity normalized
bound state $\left|\Psi_{g}\right>$  of the ground electronic potential. Like in the case of the light
induced width, the light induced shift is linearly dependent on the laser intensity $I$ and of course on
$\varepsilon_{r}$.

The photoassociation process can be viewed as a case of the optical Feshbach resonance
\cite{bohn99,fedichev96,bohn97}. The optical coupling of the ground scattering
state to the excited bound state changes both the amplitude and the phase of the scattering
wave function.  The amplitude can be detected through loss of atoms from the trap.
The phase change can also be detected \cite{fatemi00,enomoto08ofr}.
This approach leads to conclusion that the scattering length
describing ultra cold collisions of the two ground state atoms can be modified by light due
to the coupling with an excited bound state.
This effect was demonstrated with a Bose-Einstein condensate (BEC) of $^{87}$Rb \cite{theis04,thalhammer05}.
The expression for a scattering length modified by light can be written in the following form \cite{bohn99}:
\begin{widetext}
\begin{equation}
\label{equ5}
a(\Delta,I)=a_{\rm bg}+\sum_{e} l_{eg}^{\rm opt}(I,0)
\frac{\Gamma_{pe}[\Delta - \Delta_{e}-E_{e}(I,0)-E_{\rm rec,mol}]}
{[\Delta - \Delta_{e}-E_{e}(I,0)-E_{\rm rec,mol}]^{2}+[\Gamma_{e}(I,0)/2]^{2}}
\,,
\end{equation}
\end{widetext}
where $a_{\rm bg}$ is the background scattering length of the system in absence of the light.
The optical length $l_{eg}^{\rm opt}(I,\varepsilon_r)$ \cite{bohn97,ciurylo05} is
\begin{equation}
\label{equ6}
l_{eg}^{\rm opt}(I,\varepsilon_r)=\frac{\Gamma_{eg}(I,\varepsilon_{r})}{2k_{r}\Gamma_{pe}} \,.
\end{equation}
Because of the $s$-wave threshold law, this quantity approaches an energy-independent
constant as $\varepsilon_r \to 0$ that varies linearly with light intensity $I$.  The optical length
characterizes the strength of an optical Feshbach resonance, namely the ability of light to change
the scattering length of the ground state atoms; see also Ref.~\cite{chin08} for a review of Feshbach
resonances, including optical ones.  The scattering length can be changed on the order of its background
magnitude $|a_{\rm bg}|$ while minimizing losses if the optical length is very large compared
with $|a_{\rm bg}|$ so that the optical control can be achieved at large detuning.

\section{Derivation of a single channel model}\label{SingChan}

The particular properties of the photoassociation resonances are
strongly dependent on the properties of the atomic interaction of the colliding atoms.
In general we describe the colliding system using the Hamiltonian operator
$H=T+H_{A}+V_{\rm int}+V_{\rm rot}$. In this expression $T$ is the kinetic energy operator
for relative radial motion, $H_{A}$ is the atomic Hamiltonian operator representing internal
atomic degrees of freedom, $V_{\rm int}$ is the interaction operator described by nonrelativistic
molecular Born-Oppenheimer potentials, and $V_{\rm rot}$ is the rotational energy operator.
Reference \cite{ciurylo04} gives more details for the case investigated here.

Let us first focus on the interaction in the excited state. In this paper we limit our discussion
to photoassociation near the intercombination transition $^{1}S_{0}$--$^{3}P_{1}$. Therefore
we will discuss the atomic interaction properties only near the dissociation limit
$^{1}S_{0}+^{3}P_{1}$. In this case it is convenient to use the $|jlJM;p\rangle$ basis. Here
$\vec{j}$ is the total electron angular momentum, $\vec{l}$ is the rotational angular momentum,
and $\vec{J}=\vec{j}+\vec{l}$ is the total angular momentum.
The projections of $\vec{J}$ on a space-fixed $z$ axis is $M$. Finally $p$ is the total parity.
We add an index $e$ to all quantities introduced here to indicate that they correspond to the
excited electronic state. It should be noted that $J_{e}$ as well as $M_{e}$ are good quantum
numbers of the Hamiltonian $H$. In our case with $j_{e}=1$ and $p_{e}=-1$
we can solve the Schr\"odinger equation using only two channels for each $J_{e}$:
one with $l_{e}=J_{e}-1$ and one with $l_{e}=J_{e}+1$. In this basis the interaction operator
$V_{\rm int}$ is not diagonal. Its matrix elements can be expressed in terms
of potentials $V_{0}(r)$ and $V_{1}(r)$ which correspond to states with $\Omega=0$ and $\Omega=1$,
respectively, where $\Omega$ is the projection of the total electron angular momentum $j_{e}$ along
the interatomic axis. The other components of the Hamiltonian operator, $T$, $H_{A}$, and $V_{\rm rot}$,
are diagonal in our basis.

The matrix elements of the sum $V_{\rm int}+V_{\rm rot}$ can be written in a compact form
\cite{julienne86}:
\begin{widetext}
\begin{equation}
\label{equ7}
\begin{array}{c}
 \begin{array}{cc}
       l_{e}=J_{e}-1 \hspace{5cm} & l_{e}=J_{e}+1 \vspace{3mm}
 \end{array}\\
 \left(
 \begin{array}{cc}
      \frac{J_{e}}{2J_{e}+1}V_{0}(r)+\frac{J_{e}+1}{2J_{e}+1}V_{1}(r)+B(r)J_{e}(J_{e}-1) & \frac{\sqrt{J_{e}(J_{e}+1)}}{2J_{e}+1}[V_{1}(r)-V_{0}(r)]  \\
      \frac{\sqrt{J_{e}(J_{e}+1)}}{2J_{e}+1}[V_{1}(r)-V_{0}(r)] & \frac{J_{e}+1}{2J_{e}+1}V_{0}(r)+\frac{J_{e}}{2J_{e}+1}V_{1}(r)+B(r)(J_{e}+2)(J_{e}+1) \\
 \end{array}
 \right)
\end{array}
\begin{array}{c}
  \vspace{3mm} \\
  \vspace{3mm} l_{e}=J_{e}-1 \vspace{3mm}\\
  l_{e}=J_{e}+1
\end{array}
\end{equation}
\end{widetext}
where $B(r)=\hbar^{2}/(2\mu r^{2})$.  Henceforth we will omit explicit indication of the $r$-dependence
of $V_{0}$, $V_{1}$ and $B$.  Analytic diagonalization of this matrix gives the following eigenvalues:
\begin{widetext}
\begin{eqnarray}
\label{equ8}
{\cal V}_{0} &=& \frac{1}{2}\left\{
V_{0}+V_{1}-(V_{1}-V_{0})\sqrt{1-\frac{4B}{V_{1}-V_{0}}+\frac{4B^{2}(2J_{e}+1)^{2}}{(V_{1}-V_{0})^{2}}}+2B[J_{e}(J_{e}+1)+1]
\right\} \,, \\
\label{equ9}
{\cal V}_{1} &=& \frac{1}{2}\left\{
V_{0}+V_{1}+(V_{1}-V_{0})\sqrt{1-\frac{4B}{V_{1}-V_{0}}+\frac{4B^{2}(2J_{e}+1)^{2}}{(V_{1}-V_{0})^{2}}}+2B[J_{e}(J_{e}+1)+1]
\right\} \,.
\end{eqnarray}
\end{widetext}
These eigenvalues ${\cal V}_{0}$ and ${\cal V}_{1}$ can be used as effective potentials
for single channel calculations of excited bound states. Such an approach allows one to approximate
the influence of Coriolis coupling on the molecular structure in simple single channel
calculations.

The quantum number $\Omega$ becomes a good one in the range of $r$ where
$\left|V_{1}-V_{0}\right | \gg B$, i e., when the anisotropy of the interaction between two atoms
is much bigger than the rotation energy.
In such a case the eigenvalues and corresponding eigenvectors of the matrix in Eq. (\ref{equ7})
can be written in the following form:
\begin{widetext}
\begin{eqnarray}
\label{equ10}
{\cal V}_{0} &=& V_{0}+B[J_{e}(J_{e}+1)+2] \longrightarrow
\left(
\begin{array}{c}
  -\sqrt{J_{e}/(2J_{e}+1)}\vspace{3mm}\\
  \sqrt{(J_{e}+1)/(2J_{e}+1)}
\end{array}
\right) \,,\\
\label{equ11}
{\cal V}_{1} &=& V_{1}+BJ_{e}(J_{e}+1) \hspace{8mm} \longrightarrow
\left(
\begin{array}{c}
  \sqrt{(J_{e}+1)/(2J_{e}+1)}\vspace{3mm}\\
  \sqrt{J_{e}/(2J_{e}+1)}
\end{array}
\right) \,.
\end{eqnarray}
\end{widetext}

In many cases these potentials provide a successful approximation for single channel calculations
of bound states of real systems.  Such an approach is particularly good for Yb, because the potential curves
at long range are dominated by the resonant dipole interaction.
In the case of Yb $\left|V_{1}-V_{0}\right | = \frac{3}{2}C_{3}/r^{3}\gg B$ up to very large $r$.
This is a quite different case from Sr \cite{zelevinsky06}, where the $C_3$ value is more than 20 times
smaller than for Yb.   In the case of Sr the competition between the van der Waals and resonance interactions
is important. In the range of $r$ where the interaction is dominated by the resonant dipole term
the ungerade potentials for $\Omega=0$ and $\Omega=1$ are attractive and repulsive, respectively.
Therefore only the potential ${\cal V}_{0}$ will support a series of bound states having $0_{u}^{+}$
symmetry near the dissociation limit.
The ${\cal V}_{1}$ potential becomes attractive at small $r$ due to chemical bonding and may support a bound
state near the dissociation limit. In the rest of the paper we will describe excited bound states
of $0_{u}^{+}$ symmetry using single channel solutions of the Schr\"odinger equation.

This simple approach to the description of atomic interactions allows us to write
the appropriate expressions for the light induced width and shift:
\begin{equation}
\label{equ12}
\Gamma_{eg}(I,\varepsilon_{r})=\Gamma_{A}\frac{3}{4\pi}\frac{I\lambda_{A}^{3}}{c}
f_{eg}^{\rm rot} f_{eg}^{\rm FC-\Gamma}(\varepsilon_{r})
\end{equation}
and
\begin{equation}
\label{equ13}
E_{eg}(I,\varepsilon_{r})=\Gamma_{A}\frac{3}{4\pi}\frac{I\lambda_{A}^{3}}{c}
f_{eg}^{\rm rot} f_{eg}^{\rm FC-E}(\varepsilon_{r}) \,.
\end{equation}
In Eqs. (\ref{equ12}) and (\ref{equ13}) $\Gamma_{A}$ is the natural decay width of the atomic transition
and $\lambda_{A}$ the wavelength of the atomic transition.
The dimensionless rotational line strength factor for a transition from the ground scattering state to the
excited $0_{u}^{+}$ state in a bosonic isotope has the simple approximate form obtained
by Machholm {\it et al}. \cite{machholm01}:
\begin{equation}
\label{equ14}
f_{eg}^{\rm rot}=\left\{
 \begin{array}{cc}
      \frac{1}{3}\frac{2J_{e}+1}{2J_{g}+1}\frac{J_{e}}{2J_{e}+1}=\frac{1}{3}\frac{J_{g}+1}{2J_{g}+1}& \mbox{for $J_{e}=J_{g}+1$} \vspace{3mm} \,,\\
      \frac{1}{3}\frac{2J_{e}+1}{2J_{g}+1}\frac{J_{e}+1}{2J_{e}+1}=\frac{1}{3}\frac{J_{g}}{2J_{g}+1}& \mbox{for $J_{e}=J_{g}-1$} \,.
 \end{array}
\right.
\end{equation}
for even $J_{g}$.
This expression can be obtained using the approach described in Ref. \cite{ciurylo04}.
To extract the rotational line strength factor $f_{eg}^{\rm rot}$ one can start from Eq. (\ref{equ3})
and use Eq. (B3) from Ref. \cite{ciurylo04}. In our case, this allows us to show that the laser radiation
couples the ground scattering channel having $l_g=J_{g}$ only with the excited state channel having
the same $l_{e}$.
Therefore the light induced width, Eq. (\ref{equ3}), is proportional to the fraction of
the total wave function of the excited bound state in the channel with $l_{e}=l_{g}$.
This fraction can be found from Eqs. (\ref{equ10}) and (\ref{equ11}).
Moreover, the light induce width is proportional to the square of the proper Clebsch-Gordan coefficient
$\left|\left<J_{g}KM_{g}q|J_{e}M_{e}\right>\right|$, where $K=1$ for the dipole transition and
$q=-1,\; 0,\;{\rm or}\; +1$ describes the polarization of light.
This way one can introduce the rotational line strength factor
\begin{equation}
\label{equ15}
f_{eg}^{\rm rot}(M_{g},q)=\frac{2J_{g}+1}{2J_{e}+1}\left|\left<J_{g}100|J_{e}0\right>\right|^2
\left|\left<J_{g}KM_{g}q|J_{e}M_{e}\right>\right|^2,
\end{equation}
which is dependent on the light polarization $q$ and the projection $M_{g}$ of the total angular
momentum $J_{g}$ in the ground electronic state, see Ref. \cite{vogt07}.

The rotational line strength factor given by Eq. (\ref{equ14}) can be obtained as an average
\begin{equation}
\label{equ16}
f_{eg}^{\rm rot}=\frac{1}{2J_{g}+1}\sum_{M_{g}} f_{eg}^{\rm rot}(M_{g},q)
\end{equation}
over all possible orientations of the total angular momentum $J_{g}$; compare Refs. \cite{machholm01,vogt07}.
This quantity is independent of the light polarization $q$.
To derive Eq. (\ref{equ14}) one can use the following relation
\begin{equation}
\label{equ17}
\sum_{M_{g}} \left|\left<J_{g}KM_{g}q|J_{e}M_{e}\right>\right|^2=\frac{2J_{e}+1}{3}\;,
\end{equation}
which is fulfilled for $J_{e}=J_{g}-1\;{\rm and}\; J_{g}+1$, $K=1$ and $M_{e}=M_{g}+q$.
Equation (\ref{equ14}) can be also expressed in terms of a Wigner 3-$j$
symbol, see Ref. \cite{vogt07}:
\begin{equation}
\label{equ18}
f_{eg}^{\rm rot}=\frac{\left|\left<J_{g}100|J_{e}0\right>\right|^2}{3}=
\frac{2J_{e}+1}{3}\left(
\begin{array}{ccc}
J_{g}&1&J_{e}\\
0&0&0
\end{array}
\right)^{2}.
\end{equation}
This work is limited only to the case of weak interaction with light. Therefore we
can ignore the dependence of the light induced width and shift on the projection $M_{g}$ and
use the projection independent Eq. (\ref{equ14}).

The total angular momentum $J_{g}$ in the ground scattering state is the same as the rotational
angular momentum $l_{g}$. This is a consequence of the fact that the atomic interaction in the
ground electronic state has $^{1}\Sigma_{g}^{+}$ symmetry and the total electronic angular momentum
$j_{g}=0$. In homonuclear pairs of bosonic atoms only even $J_{g}$ are allowed.
Finally, the Franck-Condon factors \cite{bohn99}:
\begin{equation}
\label{equ19}
f_{eg}^{\rm FC-\Gamma}(\varepsilon_{r})=\left|\int_{0}^{\infty}dr\;\phi_{e}(r)f_{g}(r;\varepsilon_{r})\right|^{2}
\end{equation}
and
\begin{equation}
\label{equ20}
f_{eg}^{\rm FC-E}(\varepsilon_{r})=\int_{0}^{\infty}dr\;\phi_{e}(r)g_{g}(r;\varepsilon_{r})
\int_{0}^{r}dr'\phi_{e}(r')f_{g}(r';\varepsilon_{r})
\end{equation}
can be expressed in terms of the regular $f_{g}(r;\varepsilon_{r})$ and irregular $g_{g}(r;\varepsilon_{r})$
solutions of the the Schr\"odinger equation for the ground scattering state and the wave function $\phi_{e}(r)$
for the excited bound state.

The expressions for the light induced width and shift can be further simplified by using the reflection
approximation \cite{bohn99,julienne96,boisseau00b}. The Franck-Condon factors can then be written
in the following form:
\begin{equation}
\label{equ21}
f_{eg}^{\rm FC-\Gamma}(\varepsilon_{r})=\frac{\partial E_{e}}{\partial n}\frac{1}{D_{C}}
\left|f_{g}(r_{C};\varepsilon_{r})\right|^{2}
\end{equation}
and
\begin{equation}
\label{equ22}
f_{eg}^{\rm FC-E}(\varepsilon_{r})=\frac{1}{2}\frac{\partial E_{e}}{\partial n}\frac{1}{D_{C}}
g_{g}(r_{C};\varepsilon_{r})f_{g}(r_{C};\varepsilon_{r}) \,.
\end{equation}
where
$D_{C}=\left.\partial {\cal V}_{e}/\partial r\right|_{r_{C}}-\left.\partial {\cal V}_{g}/\partial r\right|_{r_{C}}$
the fraction $\frac{\partial E_{e}}{\partial n}$ is the mean vibrational spacing, ${\cal V}_{e}$ and
${\cal V}_{g}$ are the effective potentials in the excited and ground electronic states. Here $r_{C}$ is
the Condon point. When the reflection approximation is applicable, the Condon point is approximately the
classical outer turning point for the excited bound state.

The reflection approximation is very good for a number of Yb bound states close
to the $^{1}S_{0}+^{3}P_{1}$ dissociation limit.  This is in contrast
with other species such as Ca or Sr, for which the reflection
approximation needs to be modified \cite{ciurylo06}.  Since Eqs. (\ref{equ12}) and (\ref{equ21})
show that the light induced width is proportional to $|f_g(r_C)|^2$, photoassociation spectroscopy
provides an excellent tool to investigate the properties of the ground state
scattering wave function \cite{jones06}. This will be explored below for photoassociation of Yb near
the intercombination line.

\section{Experimental data}

Experimental information is crucial for determining the parameters that describe the
long range interaction near the $^{1}$S$_{0}$--$^{3}$P$_{1}$ dissociation limit.
We used the experimental data obtained by Tojo {\it et al}. \cite{tojo06} to find the parameters needed
to describe $V_0(r)$ for the $0_{u}^{+}$ state.
As described in Ref.~\cite{tojo06}, Yb atoms were decelerated by a Zeeman-slowing laser for the
$^{1}$S$_{0}$--$^{1}$P$_{1}$ transition, and were collected in a magneto-optical trap (MOT) with a laser for
the $^{1}$S$_{0}$--$^{3}$P$_{1}$ transition.
After the compression of the MOT, the atoms were transferred into a crossed optical trap with laser beams at 532 nm.
The atoms were evaporatively cooled by decreasing the trap potential depth.
Typically, MOT duration time is 10 s and evaporative cooling time is 6 s.
A laser beam for photoassociation was applied to the trapped atoms, and the number of remaining atoms was
measured through an absorption image with the $^{1}$S$_{0}$--$^{1}$P$_{1}$ transition after the release
from the optical trap.
The applied photoassociation light intensity varied between 6.5 $\mu$W/cm$^{2}$
to 90 mW/cm$^{2}$, depending on which excited level was being probed.
We obtained the atom-loss spectra by scanning the photoassociation laser frequency.
These spectra allowed the determination of the bound states energies of a
series of $J_{e}=1$ and $J_{e}=3$ $0_{u}^{+}$ levels for each of the two isotopic species.
Table \ref{tab1} lists the measured energies along with their error bars.
The data for the $^{174}$Yb$_2$ $J_{e}=1$ states were taken at about 4 $\mu $K, and the other data
in Table \ref{tab1} were taken at temperatures in the range from 5 to 27 $\mu $K.
The shift with temperature and light intensity of the photoassociation
features was taken into account in the data analysis. These shifts are mostly responsible for the magnitude
of the error bars listed in Table \ref{tab1}.

\begin{table*}
\caption{Comparison of measured binding energies in excited $0_u^+$ state for $^{174}{\rm Yb}_{2}$
and $^{176}{\rm Yb}_{2}$ with binding energies calculated from the optimal model obtained by least
squares fit of experimental data for $^{174}{\rm Yb}_{2}$ and $^{176}{\rm Yb}_{2}$ isotopes.
See details in the text. All quantities are given in MHz.}
\label{tab1}
\begin{tabular}{rrr@{\hspace{5mm}}rrr@{\hspace{5mm}}rrr@{\hspace{5mm}}rrr}
\hline
\hline
\multicolumn{6}{c}{$^{174}$Yb} & \multicolumn{6}{c}{$^{176}$Yb} \\
\multicolumn{3}{c}{$J_e=1$} &\multicolumn{3}{c}{$J_e=3$} & \multicolumn{3}{c}{$J_e=1$} & \multicolumn{3}{c}{$J_e=3$} \\
Experiment & Theory & Diff. & Experiment & Theory & Diff. & Experiment & Theory & Diff. & Experiment & Theory & Diff. \\
\hline
-4.4(1.0) & -4.2 & -0.2 &  & -3 &  &  & -3.1 &  &  & -2.1 & \\
-9.6(1.0) & -9.7 & 0.1 &  & -7.5 &  & -7.9(2.0) & -7.5 & -0.4 & -5.9(2.0) & -5.7 & -0.2 \\
-19.7(1.0) & -20.1 & 0.4 & -16.1(2.0) & -16.7 & 0.6 & -16.5(2.0) & -16.0 & -0.5 & -13.9(2.0) & -13.1 & -0.8 \\
-37.4(1.0) & -38.5 & 1.1 & -32.5(2.0) & -33.3 & 0.8 & -32.0(2.0) & -31.2 & -0.8 & -27.4(2.0) & -26.8 & -0.6 \\
-68.5(1.0) & -69.1 & 0.6 &  & -61.7 &  & -56.2(2.0) & -57.0 & 0.8 & -50.0(2.0) & -50.5 & 0.5 \\
-119.1(2.0) & -117.8 & -1.3 &  & -107.5 &  &  & -98.5 &  &  & -89.4 &  \\
-191.3(1.0) & -192.3 & 1.0 & -179.0(2.0) & -178.2 & -0.8 &  & -162.6 &  &  & -150.1 & \\
-302.3(1.0) & -302.5 & 0.2 & -284.5(2.0) & -283.9 & -0.6 & -258.2(2.0) & -258.3 & 0.1 & -242.8(2.0) & -241.7 & -1.1 \\
-461.1(1.0) & -461.3 & 0.2 &  & -437.4 &  & -398.1(2.0) & -397.1 & -1.0 & -376.4(2.0) & -375.5 & -0.9 \\
-684.6(1.0) & -684.9 & 0.3 & -654.7(2.0) & -654.6 & -0.1 & -594.4(2.0) & -593.8 & -0.6 & -567.3(2.0) & -566.3 & -1.0 \\
-993.7(1.0) & -993.7 & 0.0 &  & -955.9 &  & -867.6(2.0) & -866.7 & -0.9 & -832.7(2.0) & -832.2 & -0.5 \\
-1412.8(1.0) & -1413.0 & 0.2 & -1366.7(2.0) & -1366.4 & -0.3 & -1240.1(2.0) & -1238.9 & -1.2 &  & -1196.2 & \\
-1973.5(1.0) & -1973.9 & 0.4 & -1918.1(2.0) & -1917.3 & -0.8 &  & -1738.5 &  &  & -1686.5 & \\
\hline
\hline
\end{tabular}
\end{table*}

The same technique was used to determine  bound states energies of the excited homonuclear molecules
made from two other bosonic isotopes, $^{170}$Yb and $^{172}$Yb.
Long MOT time of about 60 s was needed for $^{170}$Yb because of its small natural abundance
\cite{fukuhara07a}, and the evaporative cooling had to be done in a short time of about 1.5 s for $^{172}$Yb
because of three-body recombination atom loss due to its large negative scattering length \cite{enomoto08ofr}.
Table \ref{tab2} lists the values of
measured binding energies in the excited $0_u^+$ $J_{e}=1$ levels of $^{170}{\rm Yb}_{2}$ and
$^{172}{\rm Yb}_{2}$ molecules.
These data were taken at about 2 $\mu $K.
The uncertainties shown in Table II are mostly due to the light-induced shift.

\begin{table}
\caption{Comparison of measured binding energies in excited $0_u^+$ state for $^{170}{\rm Yb}_{2}$
and $^{172}{\rm Yb}_{2}$ with binding energies calculated from the optimal model obtained by least
squares fit of experimental data for $^{174}{\rm Yb}_{2}$ and $^{176}{\rm Yb}_{2}$ isotopes.
See details in the text. All quantities are given in MHz.}
\label{tab2}
\begin{tabular}{rrr@{\hspace{5mm}}rrr}
\hline
\hline
\multicolumn{3}{c}{$^{170}$Yb} & \multicolumn{3}{c}{$^{172}$Yb} \\
\multicolumn{3}{c}{$J_e=1$} & \multicolumn{3}{c}{$J_e=1$}  \\
Experiment & Theory & Diff. & Experiment & Theory & Diff.  \\
\hline
 & -2.9 &  &  & -2.2 &  \\
 & -7.2 &  &  & -5.5 &  \\
 & -15.6 &  &  & -12.4 & \\
 & -31.0 &  & -26.7(1.0) & -25.1 & -1.6 \\
 & -57.3 &  & -48.9(1.0) & -47.1 & -1.8 \\
 & -99.9 &  &  & -83.3 &  \\
 & -166.1 &  & -142.6(5.0) & -140.2 & -2.4 \\
-268.1(3.0) & -265.4 & -2.7 & -228.7(1.0) & -226.4 & -2.3 \\
-412.9(3.0) & -410.4 & -2.5 & -355.3(1.0) & -352.9 & -2.4 \\
-619.3(3.0) & -616.6 & -2.7 & -544.1(5.0) & -534.1 & -10 \\
-906.3(3.0) & -903.9 & -2.4 & -798.1(5.0) & -787.9 & -10.2 \\
 & -1296.9 &  & -1145.1(5.0) & -1136.6 & -8.5 \\
 & -1826.5 &  &  & -1608.1 & \\
\hline
\hline
\end{tabular}
\end{table}

\section{Model potentials}

The location of a photoassociation resonance as the laser frequency is scanned is directly related to
the binding energy of the excited molecule. {\it Ab initio} potential curves are still
insufficiently accurate to describe the position of the most weakly bound states.
Therefore we introduce an analytic model potential valid at large $r$ and fit its parameters
to match measured bound state energies  near the dissociation limit for different isotopes.
The basic concept is similar to that used very successfully for bound state and scattering
properties of the ground state \cite{kitagawa08,vankempen02}.
The potential is chosen to have the correct long range form and an arbitrary short range form that
permits us to represent the correct absolute phase due to the unknown short range interactions.
This allows us to successfully mass scale the excited state binding energies for different isotopic combinations.
Mass scaling is more fully described in the next Section.

The interaction potential
of two atoms in the $0_{u}^{+}$ excited electronic state
for large interatomic separations can be well approximated by the following
expression (see Ref.~\cite{zelevinsky06}):
\begin{equation}
\label{equ23}
V_{0}(r) =
\frac{C^{(e)}_{6}}{r^{6}}
\left[\left(\frac{\sigma^{(e)}}{r}\right)^{6} - 1\right]
- \frac{C^{(e)}_{8}}{r^8} - \frac{C^{(e)}_{3}}{r^3} \,,
\end{equation}
where the resonant dipole-dipole interaction coefficient
\begin{equation}
\label{equ24}
C^{(e)}_{3} = \frac{3}{2} \frac{\hbar}{\tau}\left(\frac{\lambda}{2\pi}\right)^3 \,,
\end{equation}
and $\sigma^{(e)}$ is a free parameter that allows us to adjust the phase associated with the short range
potential. The wavelength of light  in vacuum corresponding to
the transition between atomic states $^{1}S_{0}$ and $^{3}P_{1}$ of Yb is $\lambda=555.802\;{\rm nm}$,
and $\tau$ is the atomic lifetime.
We use the least-squares method to optimize the values of $C^{(e)}_{3}$, $C^{(e)}_{6}$,  $C^{(e)}_{8}$,
and $\sigma^{(e)}$ while matching the calculated binding energies to the experimental values measured by
Tojo {\it et al}. \cite{tojo06} for $^{174}{\rm Yb}$ and $^{176}{\rm Yb}$.
We obtain $C_{3}^{(e)}=0.1949(11)\;E_{h}a_{0}^{3}$ which corresponds to $\tau=869.6(4.5)\;{\rm ns}$,
$C^{(e)}_{6}=2.41(0.22)\times 10^3\;E_{h}a_{0}^{6}$, $C^{(e)}_{8}=2.3(1.6)\times 10^5 \;E_{h}a_{0}^{8}$, and
$\sigma^{(e)}=8.5(1.0)\; a_0$, where $a_{0}\approx0.05292\;{\rm nm}$ and $E_{h}\approx4.360\times10^{-18}\;{\rm J}$.
The errors quoted give the one standard deviation statistical fitting error,
and do not reflect any systematic errors in the model. Adding the $C^{(e)}_{8}$ coefficient to the model was
needed to improve the quality of the fit for $J_{e}=3$ levels. Introducing $C^{(e)}_{8}$ allowed us to determine
the sensitivity of $C^{(e)}_{6}$ to variation of the shorter range of the potential. This sensitivity also
contributes to the standard deviation of $C^{(e)}_{6}$. For completeness we give the fitting parameters
to enough significant digits to reproduce the calculated values to 10 kHz, as shown in Table \ref{tab1}:
$C^{(e)}_{3}=0.19488626 \;E_{h}a_{0}^{3}$ ($\tau=869.64762\;{\rm ns}$),
$C^{(e)}_{6}=2405.3647 \;E_{h}a_{0}^{6}$,
$C^{(e)}_{8}=229451.31 \;E_{h}a_{0}^{8}$, and
$\sigma^{(e)}=8.4897163 \; a_0$.

Table \ref{tab1} compares the binding energies predicted by the model  to the experimental data for
$^{174}\rm{Yb}_2$ and $^{176}\rm{Yb}_2$. The model predicts the experimental values to within about one MHz on
the average, consistent with the experimental error bars.

Having the data for two different isotopes allows us to construct an appropriately mass-scaled model for the excited
state levels, similar to what is possible for the ground state~\cite{kitagawa08}.  While cautioning that the short
range physics is complicated by the interaction of other molecular states with the $0_{u}^{+}$ state and may not be
fully represented by a single potential, our mass scaled model determines that the most deeply bound observed
levels at $-1973.5$ MHz for $^{174}\rm{Yb}_2$ and $-1240$ MHz for $^{176}\rm{Yb}_2$ respectively correspond to
the $v=$ 106 and 108 vibrational levels of the model potential.

An excellent test of mass scaling is to test it using other isotopic combinations.
Table \ref{tab2} shows a comparison of our new measured binding energies for $^{170}\rm{Yb}_2$ and $^{172}\rm{Yb}_2$
with those predicted by our mass scaled single-potential model.  There is reasonable agreement of about
3 MHz between measured and calculated levels, except for the three most deeply bound levels observed
for $^{172}\rm{Yb}_2$. While 3 MHz is on the order of the experimental uncertainty,
the accuracy of mass scaling may be more limited
for the excited states than for the ground states, where it was found to be good to approximately
0.1 MHz~\cite{kitagawa08} for binding energies up to 325 MHz.
Another source of error in our single channel excited state model could be the neglect of interactions
with short range eigenstates of other symmetries.  For example, it may be that bound states of $1_{u}$
symmetry near the threshold perturb the $0_{u}^{+}$ bound states that are near in energy to them. This
could be the reason for relatively large deviations for the three most deeply bound states of $^{172}\rm{Yb}$.

Our model potential describing the interaction in the electronic excited state $0_{u}^{+}$
has a very different shape from those obtained in an {\it ab initio} calculation by Wang and Dolg \cite{wang98}.
Clearly such a model can not be treated as a good representation of the short range interaction.
Nevertheless, its applicability over a range of isotope masses gives us confidence that the number
of bound states determined from our model is correct to within one or two bound states.
The vibronic quantum number for  the $^{174}\rm{Yb}$ bound state at -4.4 MHz  is 118 with the ground state
labeled as zero. Also the long range interaction should be well described by the model used here.

\begin{table}
\caption{Corrections to binding energies of the $0_u^{+}$ bound states with $J_{e}=1$
in the  $^{174}{\rm Yb}_{2}$ molecule caused by the Coriolis coupling and retardation
effect; see the text for details. All calculated values are given in MHz.}
\label{tab3}
\begin{tabular}{r r r r}
\hline
\hline
Binding & \multicolumn{3}{c}{Corrections} \\
Energy  &\; Coriolis \; &\; Retardation 1\; & \;Retardation 2 \\
\hline
   -4.2 & -0.004 & -0.5  &  -0.5\\
   -9.7 & -0.005 & -0.6  &  -0.6\\
  -20.1 & -0.006 & -0.8  &  -0.7\\
  -38.5 & -0.008 & -1.0  &  -0.9\\
  -69.1 & -0.009 & -1.2  &  -1.0\\
 -117.8 & -0.011 & -1.4  &  -1.1\\
 -192.3 & -0.013 & -1.7  &  -1.2\\
 -302.5 & -0.015 & -1.9  &  -1.2\\
 -461.3 & -0.017 & -2.2  &  -1.2\\
 -684.9 & -0.020 & -2.5  &  -1.1\\
 -993.7 & -0.022 & -2.8  &  -0.9\\
-1413.0 & -0.025 & -3.1  &  -0.5\\
-1973.9 & -0.027 & -3.5  &   0.0\\
\hline
\hline
\end{tabular}
\end{table}

We have tested the possible influence of the Coriolis coupling on the results of
our calculations. We have compared our calculation used for the fits in which effective potential
was given by Eq. (\ref{equ10}) with those obtained using the effective potential in the form of
Eq. (\ref{equ8}), and find only a small difference
in binding energies for $0_{u}^{+}$ bound states with $J_{e}=1$ in the $^{174}{\rm Yb}_{2}$ molecule.
Table \ref{tab3} shows the magnitude of the Coriolis corrections in the column labeled "Coriolis".
These corrections calculated from the single channel model, Eq. (\ref{equ8}), are below
30 kHz and much less than the experimental error bars.  The relative importance of this correction increases
as the bound state energy approaches the threshold.  The Coriolis coupling mostly
play a marginal role in the Yb$_2$ excited state system.   By contrast, Ref.~ \cite{zelevinsky06} showed
it needs to be taken into account to correctly calculate two most weakly bound $0_{u}^{+}$ states
with $J_{e}=1$ in the $^{88}{\rm Sr}_{2}$ molecule.

Similar tests were carried out to check the possible influence of retardation effects \cite{power67,meath68}.
To do this we have replaced in the effective potential, Eqs. (\ref{equ10}) and (\ref{equ23}),
the standard term describing the resonance interaction $-C_{3}^{(e)}/r^{3}$ by the term in which retardation
is taken into account: $-(C_{3}^{(e)}/r^{3})[\cos(r/\lambdabar)+(r/\lambdabar)\sin(r/\lambdabar)]$,
where $\lambdabar=\lambda/(2\pi)$ \cite{meath68,machholm01}. The corrections to the binding energies caused
by this change are listed in the column "Retardation 1" of Table \ref{tab3}. These corrections are on
the order of a few MHz.  In contrast, the same corrections for $^{88}{\rm Sr}_{2}$ molecule
\cite{zelevinsky06} are more than one order of magnitude smaller.  To check whether such shifts might
be detectable in fitting the data, we have modified $\sigma^{(e)}$ so as to change the quantum defect and
fit the binding energy of the most bound level. The column labeled "Retardation 2" show the
differences from the values calculated without retardation.
Since these differences are on the order of 1 MHz or less, comparable to the experimental uncertainty,
the present data are not sufficiently accurate to come to any definitive conclusions about the observability
of retardation corrections.

\begin{table}
\caption{Comparison of experimental ground state binding energies
\cite{kitagawa08} with the present model. All values are given in
MHz.  The last column is the difference between the experimental
values and the present model.} \label{tab4}
\begin{tabular}{r r r r r r r}
\hline \hline
Isotope & v & $J_g$ & Experiment & Theory & Theory & Diff. \\
& & & \cite{kitagawa08} & \cite{kitagawa08} & Present work & \\
\hline
$^{176}$Yb & 1 & 0 & -70.404 & -70.405 & -70.378 & -0.026 \\
 & 1 & 2 & -37.142 & -37.118 & -37.093 & -0.049 \\
$^{174}$Yb & 1 & 0 & -10.612 & -10.642 & -10.629 & 0.018 \\
 & 2 & 0 & -325.607 & -325.607 & -325.602 & -0.005 \\
 & 1 & 2 & -268.575 & -268.576 & -268.570 & -0.005 \\
$^{173}$Yb & 1 & 0 & -1.539 & -1.613 & -1.609 & 0.070 \\
$^{172}$Yb & 1 & 0 & -123.269 & -123.349 & -123.321 & 0.052 \\
 & 1 & 2 & -81.786 & -81.879 & -81.851 & 0.065 \\
$^{171}$Yb & 1 & 0 & -64.418 & -64.548 & -64.522 & 0.104 \\
 & 1 & 2 & -31.302 & -31.392 & -31.367 & 0.065 \\
$^{170}$Yb & 1 & 0 & -27.661 & -27.755 & -27.735 & 0.074 \\
 & 1 & 2 & -3.651 & -3.683 & -3.667 & 0.016 \\
\hline \hline
\end{tabular}
\end{table}

Reference~\cite{kitagawa08} fits ground state binding energy data for two isotopes using a similar form for
the ground state potential without the dipolar term,
\begin{equation}
\label{equ25}
V_{g}(r) = \frac{C^{(g)}_{6}}{r^{6}}\left[\left(\frac{\sigma^{(g)}}{r}\right)^{6} - 1\right]
- \frac{C^{(g)}_{8}}{r^8} \,;
\end{equation}
This model with mass scaling
predicted the observed binding energies for 4 other isotopic molecules with an error on the order
of $0.1$ MHz or less. Here we have refitted the ground state potential by simultaneously fitting the data for
all isotopes in Ref.~\cite{kitagawa08}.  The fit has a slightly better $\chi^2$ than the previous one, but
does not represent a significant improvement. For the sake of completeness, we give the model parameters for
the global fit with enough significant digits to reproduce the calculated values in Table \ref{tab4}:
$C^{(g)}_{6}=1930.2481\;E_{h}a_{0}^{6}$,
$C^{(g)}_{8}=194683.32\;E_{h}a_{0}^{8}$ and
$\sigma^{(g)}=9.0240156\;a_{0}$.

The $C^{(g)}_{6}$ and $C^{(g)}_{8}$ obtained here as well as in Ref.~\cite{kitagawa08} agree
very nicely with those calculated by Zhang and Dalgarno \cite{zhang08}. The reported values
in Ref. \cite{zhang08} are $C^{(g)}_{6}=2070\;E_{h}a_{0}^{6}$ and
$C^{(g)}_{8}=2.023\;\times 10^{5} E_{h}a_{0}^{8}$, respectively, with an estimated uncertainty of 10\%.
It should be emphasized that the shape of our potential can significantly differ form the
real one. Nevertheless our model potential should correctly represent the number of vibronic bound states
in the ground electronic state of Yb$_{2}$ molecule. Therefore even if our model is significantly different
from an {\it ab initio} potential like that reported by Buchachenko {\it et al}. \cite{buchachenko07}
it should give about the same number of vibronic bound states.

It should be noted that similar analysis  to that for Yb \cite{kitagawa08} was carried out recently
for the scattering properties in the ground electronic state of various isotopes
of Sr by Martinez {\it et al}, using a realistic potential \cite{martinez08}.
The scattering properties from this work are in very good agreement with results obtained using a potential
derived from Fourier transform spectroscopy
by Stein {\it et al}. \cite{stein09}.

\section{Ground state scattering wave function}

As discussed in Section~\ref{SingChan}, the strength and shape of a photoassociation line is to a large
extent determined by the ground state scattering wave function at the Condon point for the transition.
Since the Condon points for photoassociation to near-threshold excited states tend to be
at quite large internuclear separation $r$, the scattering wave function needs to be known
only at relatively large $r$.  Consequently, in order to describe photoassociation in the
ultra low scattering energy regime, the detailed knowledge of the atomic interaction at
short range can be compressed to a very few parameters.

The most important quantity describing scattering during an ultracold collision is
the scattering length $a$.  If the long range potential has the van der Waals form $V_{g}(r)=-C_{6}/r^{6}$, the scattering length is very well
approximated by a simple analytical relation given by Gribakin and Flambaum \cite{gribakin93}
\begin{equation}
\label{equ26}
a = \bar{a} \left[ 1-\tan\left(\Phi-\frac{\pi}{8}\right)\right]
\end{equation}
see also Refs. \cite{flambaum99,boisseau00a}.
Here the mean scattering length
$\bar{a}=2^{-3/2}\frac{\Gamma(3/4)}{\Gamma(5/4)}\left(2\mu C_6/\hbar^2\right)^\frac{1}{4}$
is a characteristic length associated with the van der Waals potential,
where $\Gamma$ is the gamma-function. This length also defines a characteristic
energy for the van der Waals potential,
$\bar{\varepsilon}=\hbar^{2}/(2\mu\bar{a}^{2})$.
The phase $\Phi$ is defined by
\begin{equation}
\label{equ27}
\Phi = \frac{\sqrt{2\mu}}{\hbar}\int_{r_0}^{\infty} \sqrt{-V_{g}(r)}dr,
\end{equation}
where $r_0$ is the inner classical turning point of $V(r)$ at zero energy.
The number of bound states $N$ in the potential is~\cite{flambaum99}
\begin{equation}
\label{equ28}
 N=\left [ \frac{\Phi}{\pi} -\frac{5}{8} \right ] +1,
\end{equation}
where the brackets $[ \ldots ]$ mean the integer part.

The scattering length varies periodically with phase $\Phi$, having a singularity when $\Phi/\pi=N-3/8$.
This variation can be
observed experimentally for different isotopes of the same species. If we assume that the
interaction potential is the same for all isotopes so that mass scaling applies,
$\Phi(\mu)$ can be varied by changing the reduced mass $\mu$.
The relation between scattering length, the energies of near threshold bound states
and the reduced masses of the colliding atoms was carefully studied
by Kitagawa {\it et al}. \cite{kitagawa08} for ground state interactions of various Yb isotopes.
It is useful to define a reduced mass difference $\Delta\mu$ needed to change $\Phi(\mu)$ by $\pi$,
that is, to change the number of bound states in the potential by one. It is approximately
\begin{equation}
\label{equ29}
\Delta\mu \approx 2\mu/N \,.
\end{equation}
Since $N=72$ for $^{174}$Yb, we see that a mass difference of 5 atomic mass units is sufficient to change
the number of Yb$_2$ bound states by one.  Alternatively, varying $\mu$ continuously by 5 atomic mass
units will cause the scattering length to vary across a singularity over its full range from $-\infty$
to $+\infty$, as found by Ref.~\cite{kitagawa08}.  There are actually 7 stable isotopes of Yb,
and thus 28 different discrete physical values of $2\mu$ that are available in the laboratory using
different isotopic combinations.

\begin{figure}
\includegraphics[angle=0,width=\columnwidth,clip]{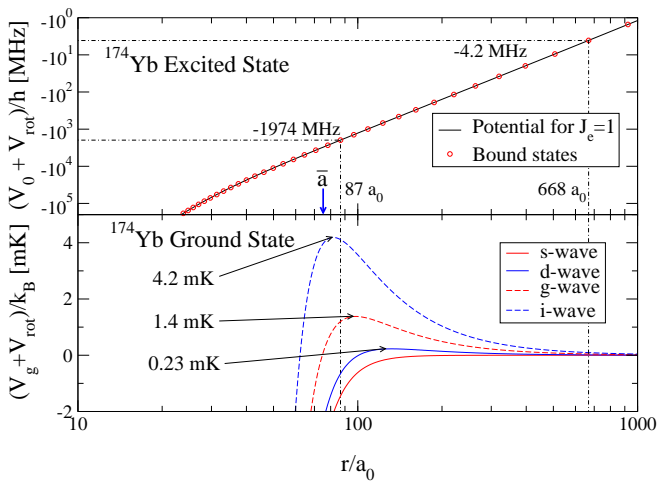}
\caption{(Color online)
The lower and upper panels show $V_{g}+V_{\rm rot}$ and $V_{0}+V_{\rm rot}$, the respective ground state
and excited state potentials.  The upper curve shows the outer turning points of each calculated
$0_u^+$, $J_{e}=1$ bound level for $^{174}$Yb.
The dotted lines indicate the experimentally observed range of levels.
The arrow indicates the mean scattering length $\bar{a}=75.18\;a_{0}$.
The lower panel shows the centrifugal barriers for the long range ground state potential
for $J_{g}=l_{g}=$ 0 ($s$-wave), 2 ($d$-wave), 4 ($g$-wave), and 6 ($i$-wave).
An energy $(V_{g}+V_{\rm rot})/k_B=$ 1 mK is equivalent to $(V_{g}+V_{\rm rot})/h=$ 21 MHz.}
\label{fig1}
\end{figure}

The long-range ground state scattering wave function at very low collision energy depends on three basic parameters, namely the separation between colliding atoms $r$, the relative kinetic energy of collision
$\varepsilon_{r}$, and the quantum defect associated with the phase $\Phi$ or, equivalently, the scattering length $a$.
The reduced mass $\mu$ is a parameter that allows control of the quantum defect. Thus, the
long-range scattering wave function for the van der Waals system has an universal character that can be expressed
by a system-independent function of the dimensionless variables $r/\bar{a}$, $\varepsilon_{r}/\bar{\varepsilon}$
and $\mu/\Delta\mu$. All of the needed information about the ground state scattering and bound state wave functions can be calculated from a knowledge of the
scaling parameters $r/\bar{a}$, $\varepsilon_{r}/\bar{\varepsilon}$
and $(\mu-\mu_{0})/\Delta\mu$ and the value of reduced mass $\mu_{0}$ for which the scattering
length is singular in a given system.

The analytical solutions of the Schr\"odinger equation are known for several class of potentials, thus allowing the derivations of compact expressions for the scattering length \cite{szmytkowski95}.
Van der Waals systems such as those discussed here can be nicely described
and very well understood using analytical theory.   Gao \cite{gao00,gao01} used the framework of quantum defect theory for a van der Waals system to work out a number of practical
formulas and results.  For example, when the $s$-wave scattering length is singular, there is a bound state at $E=0$ not only for the $s$-wave but also for
$l_{g}=4,8,12,...$.  On the other hand if the scattering length is equal to $\bar{a}$ there will be zero
energy bound states for $l_{g}=2,6,10,...$. The practical consequence for real potentials is that when
the $s$-wave scattering length is near such special values, these other threshold bound states show up
as shape resonances for other partial waves.  A shape resonance is a quasibound state with enhanced amplitude
trapped behind a centrifugal barrier that can lead to enhanced photoassociation.  While the wave functions
we show are calculated numerically by solving the Schr\"odinger equation for our potential,
the analytic results are very helpful for interpreting their features, and for evaluating approximations
like the reflection formula.

\begin{figure*}
\includegraphics[angle=0,width=2.0\columnwidth]{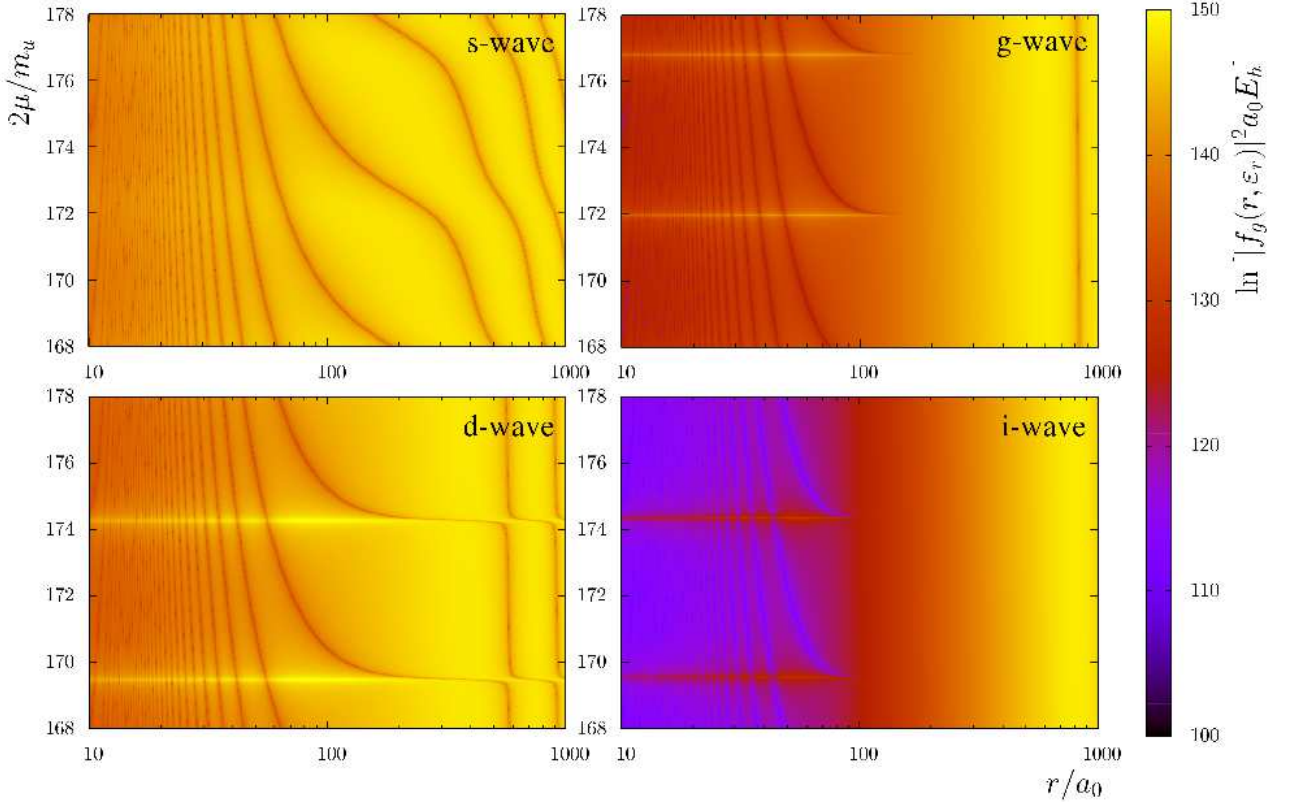}
\caption{(Color online)
The squared magnitude of the numerically calculated scattering wave function $|f_{g}(r,\varepsilon_{r})|^{2}$
of two colliding ground state Yb  atoms is shown as a function of intermolecular separation $r$ and reduced
mass of colliding system $\mu$, where $m_{u}\approx1.6605\times10^{-27}\;{\rm kg}$ .
The calculations were carried out for energy of collision $\varepsilon_{r}=100\;\mu K$
and for $J_{g}=l_{g}=$ 0 ($s$-wave), 2 ($d$-wave), 4 ($g$-wave), and 6 ($i$-wave).
The dark lines show the nodal lines of the wave functions, whereas the bright areas show the largest
amplitudes.}
\label{fig2}
\end{figure*}

As we already emphasized before, a photoassociation experiment can be used to map the
square of the scattering wave function in the ground electronic state at the Condon points corresponding
to the different excited bound levels.
Figure \ref{fig1} shows on the lower panel the effective potential,
$V_{g}(r)+B(r)J_{g}(J_{g}+1)$, in the ground electronic state, where $J_g$ is the rotational quantum number for
the partial wave $\ell = J_g$.
The Fig. \ref{fig1} shows the centrifugal barrier for a few partial waves with low $J_g$.
The upper panel of this figure shows the excited state $0_u^+$ potential  together
with bound states and their outer turning points, which are essentially the same as their Condon points.
This panel shows how the $r_C$ value that is probed changes with the binding energy of the excited bound
level. Probing experimental levels with binding energies $E_b/h$ between 2000 MHz and 3 MHz allow probing
the scattering wave function between around 80 a$_0$ and 800 a$_0$.

In order to illustrate some properties of the scattering wave function we will begin by setting the
collision energy to a constant value corresponding to $\varepsilon_{r}/k_B =100$ $\mu$K.
As we will show in the next section, shape resonances could be clearly observed at this energy,
which is below the top of the $g$-wave centrifugal barrier.   Figure \ref{fig2}
shows the dependence of the square of the wave function magnitude on $\mu$ and $r$.
Changing $\mu$ corresponds to the change of the quantum defect of the colliding system,
which changes $a$ and the bound state spectrum.  Calculations have been done for a few lowest partial
waves $s$, $d$, $g$, and $i$.

In Fig. \ref{fig2} the brighter areas show regions of higher amplitude and the dark lines show the nodes
of the wave function.  The effect of the exclusion of the wave function from short range by the centrifugal
barrier is especially evident for the $g$-wave and the $i$-wave. For this Yb system, the singularities in
the $s$-wave scattering length occur for $2\mu = $ 167.3, 172.1, and 176.9 $m_{u}$~\cite{kitagawa08}.
Near these values, an $s$-wave node moves in to smaller distances on the order of
$\bar{a}=75.18 \;a_{0}$
as $2\mu$ increases and a new bound state appears in the potential.
There is also a $g$-wave shape resonance with a dramatic enhancement of short-range amplitude near
these $2\mu$ values where $a$ is singular, as expected from the analytic van der Waals quantum defect theory.
Similarly, there are $d$- and $i$-wave shape resonances near the values of $2\mu=$ 169.7 and 174.5 $m_{u}$
where $a=\bar{a}$ for the Yb system.  Near these shape resonances, a node moves to shorter distances inside
the centrifugal barrier as $2\mu$ increases and there is an extra bound state in the potential.

If the collision energy $\varepsilon_{r}$ is lowered, the wave functions show similar patterns,
except that the short-range wave function is much more attenuated due to lower penetration through
the centrifugal barrier.  Clearly, with 28 different physical values for $2\mu$ available, we can expect
significant isotopic variation in the photoassociation spectra, depending on the range of Condon points
sampled and the temperature of the sample.    Spectral lines with Condon points outside of the barrier will
have less isotopic sensitivity, whereas lines with Condon points in or inside the region of the barrier will
show much more sensitivity to the isotopic combination.

\section{Isotopic variation of photoassociation spectra}

In this section we will show how the scattering properties of the different Yb isotopes
affect their photoassociation spectra. As an example we performed calculations of the light induced
trap-loss coefficient at various gas temperatures in range from $10\; {\rm \mu K}$, to $1\; {\rm mK}$.
Results for $T=100\; {\rm \mu K}$ are particularly interesting, since the line shapes are relatively sharp
and ground scattering resonances can have significant influence on the spectrum.
To emphasize the qualitative differences between spectra of the different isotopes we first focus
our attention on relatively deep bound states having binding energy around 2000 MHz.
As can be seen in Fig. \ref{fig1} this corresponds to the Condon point placed near or inside the centrifugal
barriers.

\begin{figure}
\includegraphics[angle=0,width=\columnwidth,clip]{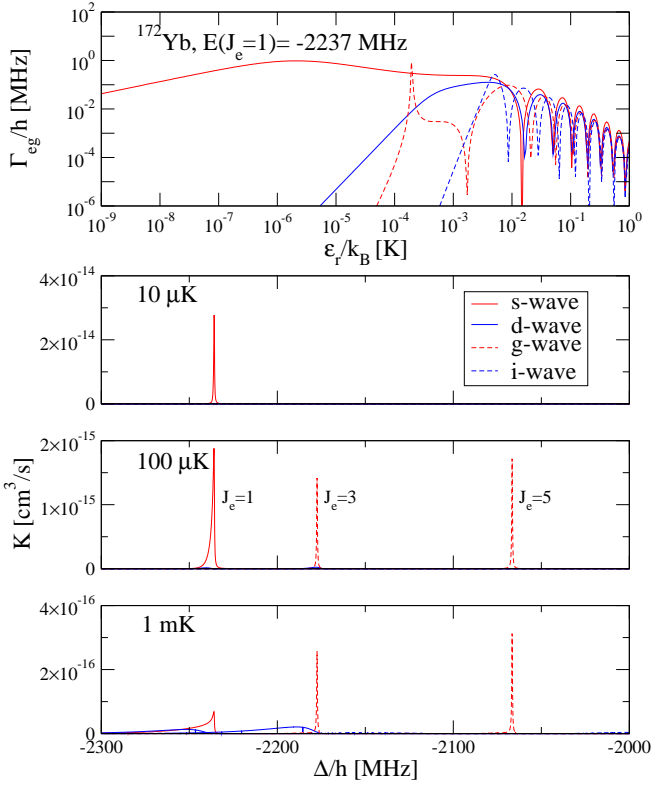}
\caption{(Color online)
Results for the excited $0_{u}^{+}$ bound states of the $^{172}$Yb$_{2}$ molecule near the $J_{e}=1$
level at $E/h=-2237\;$ MHz; other rotational levels with the same vibrational quantum number are evident.
The top panel shows contributions to the light induced width $\Gamma_{eg}$
coming from the optical coupling of the ground scattering states having
$J_{g}=l_{g}=$ 0 ($s$-wave), 2 ($d$-wave), 4 ($g$-wave), and 6 ($i$-wave)
with excited bound states having $J_{e}=J_{g}+1$, as a function of the collision energy $\varepsilon_{r}$,
calculated for the light intensity $1\;{\rm W/cm^{2}}$.
The $s$-wave scattering length is quite large in magnitude, $-598(63)\;a_{0}$~\cite{kitagawa08}, so that
this system has a $g$-wave shape resonance near $\varepsilon_{r}/k_B=200$ $\mu$K that leads to resonantly
enhanced $J_e=3$ and 5 features in the photoassociation spectrum.
The lower three panels show contributions to the fully thermally averaged loss rate $K$
coming from the optical coupling of the ground scattering states having
$J_{g}=l_{g}=$ 0 ($s$-wave), 2 ($d$-wave), 4 ($g$-wave), and 6 ($i$-wave)
with excited bound states having $J_{e}=J_{g}+1$ and $J_{e}=J_{g}-1$, as a function of the laser
detuning from the atomic resonance, calculated for various temperatures $T=10\;\mu{\rm K}$,
$100\;\mu{\rm K}$, and $1\;{\rm mK}$ and the light intensity $0.1\;{\rm mW/cm^{2}}$.}
\label{fig3}
\end{figure}

We start our discussion from $^{172}$Yb which has a large negative scattering length
near an $s$-wave singularity \cite{kitagawa08}. Consequently, we expect a $g$-wave shape resonance
at low scattering energy. The upper panel of Fig. \ref{fig3} shows the energy dependence of the light induced
width for the $J_{e}=$1, 3, 5 and 7 levels with the same vibrational quantum number
as the $E(J_{e}=1)/h=-2235\;{\rm MHz}$ level.   The light induced widths
were calculated for scattering states with total angular momentum $J_{g}=J_{e}-1$.
The widths for the same scattering partial wave $J_g$ but with $J_{g}=J_{e}+1$ will be very similar in energy
variation.  The $g$-wave resonance can be clearly seen in upper panel of Fig. \ref{fig3}.

To see how the photoassociation spectrum changes with the temperature of the gas sample we have
calculated the thermally averaged light induced trap-loss coefficients for temperatures
$T=$10 $\mu$K, 100 $\mu$K and 1 mK; see Fig. \ref{fig3}.  These correspond to $kT/h$ of 210 kHz, 2.1 MHz,
and 21 MHz respectively. At temperature 10 $\mu$K the spectrum is narrow, with a width determined mainly by
the spontaneous emission rate, and is dominated by $s$-wave scattering, for which only the $J_{e}=1$
state is visible. The calculated spectrum at temperature $T=100\; {\rm \mu K}$ has three lines.
One is the $J_{e}=1$ line due to $s$-wave scattering.
This line has a normal thermal line shape~\cite{jones06}.
Two other strong and sharp lines corresponding to $J_{e}=3$ and $J_{e}=5$ bound states
are due to the $g$-wave resonance shown in upper panel of Fig. \ref{fig3}.
Both lines have subthermal width; see Ref. \cite{machholm02}.
The shape of these two lines is determined mostly by the shape of the $g$-wave resonance.
Finally, at a temperature of 1 mK one can clearly see the very broad thermally broadened lines coming from
$s$ and $d$-wave partial waves. The $d$-wave feature is much broader than the $s$-wave one.
This is caused by the fact that for collision energy around 1 mK the light induced width
for the $s$-wave decreases with an increase of the collision energy, while for the $d$-wave the light
induced width increases. This affects the shape of the line and leads to some shifting and widening
of the line coming from the $d$-wave compared to that coming from the $s$-wave.
In addition, the $g$-wave shape resonance gives rise
to a sharp subthermal feature.

\begin{figure}
\includegraphics[angle=0,width=\columnwidth,clip]{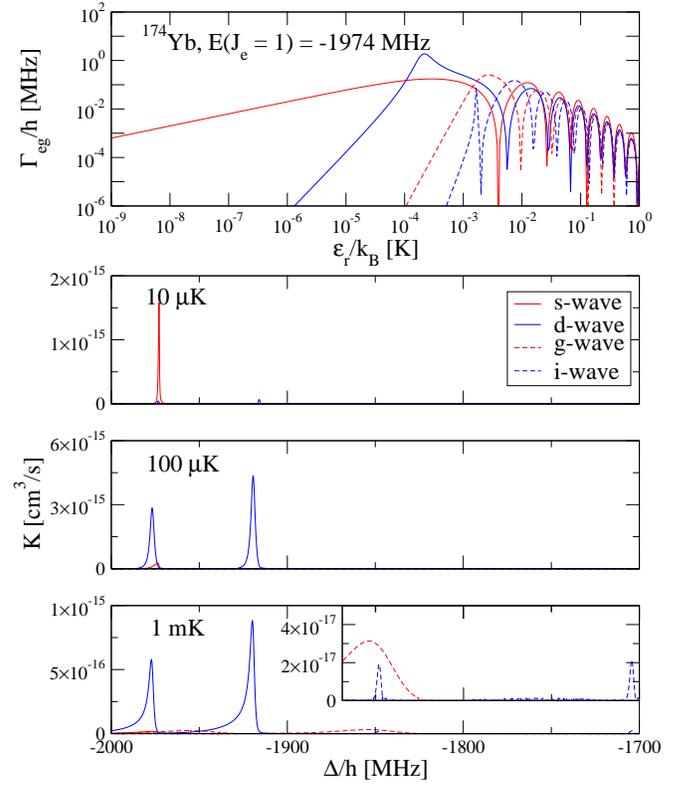}
\caption{(Color online)
Results for the excited $0_{u}^{+}$ bound state of the $^{174}$Yb$_{2}$ molecule near the $J_{e}=1$ level at
$E/h=-1974$ MHz.  The panels are the same as described in the caption of Fig \ref{fig3}.
The $s$-wave scattering length is $105(2)\;a_{0}$, which is about $1.4\bar{a}$, and there is a broad
$d$-wave shape resonance near $\varepsilon_{r}/k_B=220$ $\mu$K near the top of the $d$-wave centrifugal
barrier~\cite{kitagawa08}.}
\label{fig4}
\end{figure}

The second example is for $^{174}$Yb, which has a measured scattering
length relatively close to $\bar{a}$ \cite{kitagawa08,enomoto07}.
In such case one can expect that both a $d$-wave and an $i$-wave resonance can occur near threshold.
As with Fig. \ref{fig3} for $^{172}$Yb  Fig. \ref{fig4} shows the light induced width for the $^{174}$Yb
bound-states with the total angular momentum $J_{e}=$1, 3, 5 and 7, with the same vibronic quantum
number as the level at $E(J_{e}=1)/h=-1973$ MHz. The figure clearly shows the $d$ and $i$-wave resonances.
As for $^{172}$Yb, the spectrum at a temperature of 10 $\mu$K is dominated by the $s$-wave component.
However, unlike $^{172}$Yb, the $d$-wave shape resonance leads to a weak $J_{e}=3$
feature even at such low temperature.  The spectra at $T=100$ $\mu$K and 1 mK clearly show the influence
of the $d$-wave resonance. This resonance had a crucial role in the correct interpretation of
photoassociation spectra near the resonance transition $^{1}S_{0}-^{3}P_{1}$ and in the
determination of the scattering length for this isotope \cite{enomoto07}.
When the spectrum is dominated by the features due to the ground state $d$-wave, the intensity ratio of the
$J_{e}=3$ and $J_{e}=1$ features should be 3/2 from Eq. (\ref{equ14}).
Such $d$-wave doublets in $^{174}$Yb were observed by Tojo {\it et al} \cite{tojo06},
who found that the transition to the bound state with $J_{e}=3$ can be stronger than the transition
to $J_{e}=1$ at relatively low collisions energies corresponding to a temperature of about $25\;\mu{\rm K}$,
what can be explained by our model.

The spectrum at temperature 1 mK is also dominated by the $d$-wave component. However one can notice sharp
structure on the top of a weak and broad feature due to the $J_{e}=5$ bound-state, mostly due to the ground
state $g$-wave. The sharp structure is a consequence
of the $i$-wave resonance. The $i$-wave component has a subthermal
width connected to the $i$-wave resonance seen in upper panel of Fig. \ref{fig4}.
One can also notice a weak line corresponding to $J_{e}=7$ bound state supported by this
$i$-wave resonance.

\begin{figure}
\includegraphics[angle=0,width=\columnwidth,clip]{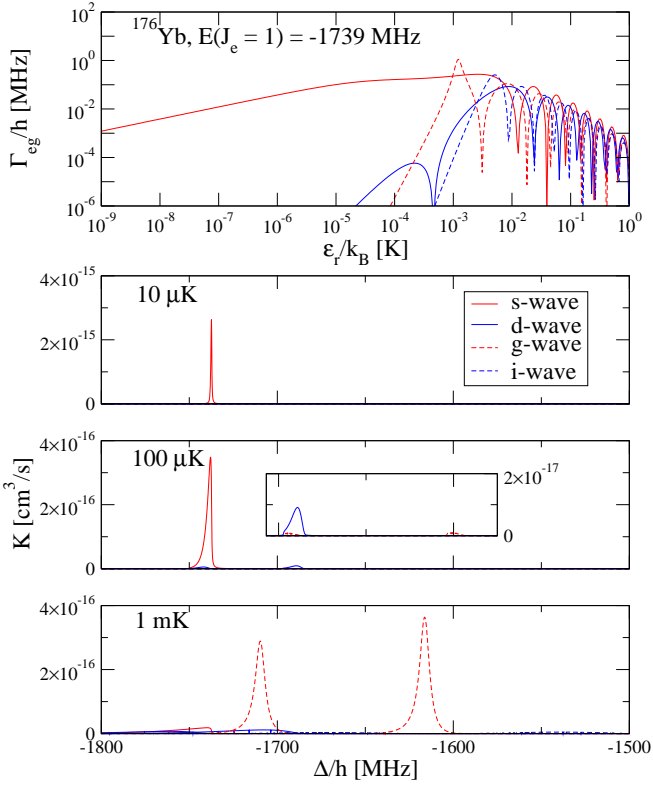}
\caption{(Color online)
Results for the excited $0_{u}^{+}$ bound state of the $^{176}$Yb$_{2}$ molecule near the $J_{e}=1$ level
at $E/h=-1739$ MHz.  The panels are the same as described in the caption of Fig \ref{fig3}.
The $s$-wave scattering length is $-24(5)\;a_{0}$, not near any special value, but there is a broad
$g$-wave shape resonance at relatively high energy near $\varepsilon_{r}/k_B=1$ mK near the top of the
$g$-wave centrifugal barrier~\cite{kitagawa08}.}
\label{fig5}
\end{figure}

Figure \ref{fig5} shows our calculations for $^{176}$Yb. This isotope has a relatively small negative
scattering length \cite{kitagawa08}.  The figure shows the light induced width for
$^{176}$Yb bound-states with the total angular momentum $J_{e}=$1, 3, 5 and 7, with the same vibronic quantum
number as the level at  $E(J_{e}=1)/h=-1737\;{\rm MHz}$.
Although the scattering length is not near any special value for van der Waals quantum defect theory,
a wide $g$-wave resonance is clearly seen relatively high in energy near 1 mK.
The spectrum at 10 $\mu$K is typical with only an $s$-wave line that can be observed.
Upon increasing the temperature to 100 $\mu$K the spectrum is still dominated by the $s$-wave component,
although noticeable $d$-wave components begin to appear.
Moreover very weak $g$-wave components occur in this spectrum.
The picture changes dramatically when the temperature increases to 1 mK, where the $g$-wave components
dominate the spectrum.

\begin{figure}
\includegraphics[angle=0,width=\columnwidth,clip]{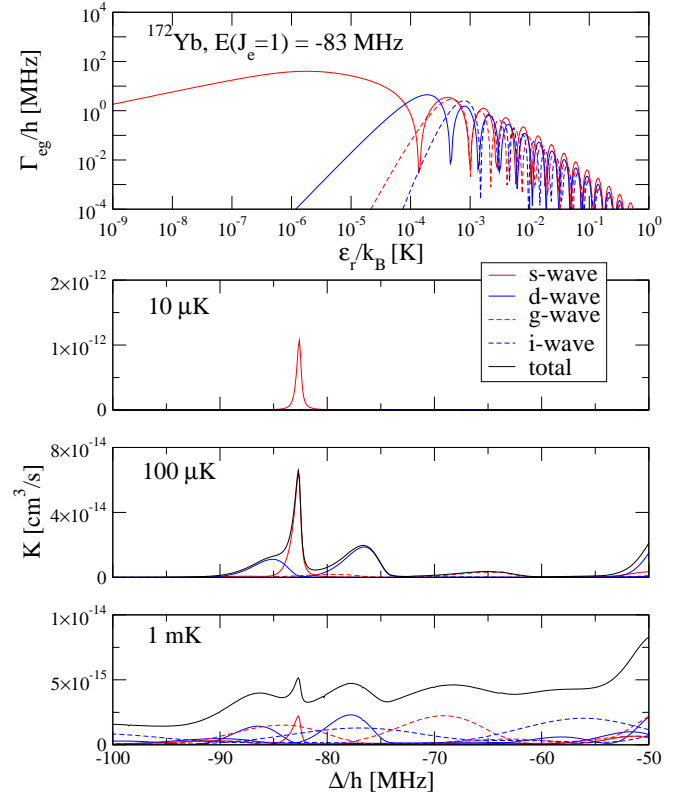}
\caption{(Color online)
Results for the excited $0_{u}^{+}$ bound state of the $^{176}$Yb$_{2}$ molecule near the $J_{e}=1$ level at $E=-83$ MHz.  The panels are the same as described in the caption of Fig \ref{fig3}.  In this case the binding energy is much smaller than in Figs. \ref{fig3}, \ref{fig4}, and \ref{fig5}, and the Condon points are outside the centrifugal barriers except when the collision energy is very low; see Fig. \ref{fig1}.}
\label{fig6}
\end{figure}

Finally we would like to show how the spectrum is affected when the Condon point for the transition is moved
outside the region of the centrifugal barrier. This is done by choosing a level with much smaller binding
energy in our previous examples.  Figure \ref{fig6} shows the light induced width for the
$^{172}$Yb bound-states with the total angular momentum $J_{e}=$1, 3, 5 and 7, with the same vibronic quantum
number as the level at  $E(J_{e}=1)/h=-83$  MHz.  The upper panel of Fig. \ref{fig6} shows that  there are
no shape resonances, since there are no amplitude enhancements from being behind the centrifugal barrier.
The lower panels of Fig. \ref{fig6} show
the calculated spectra at 10 $\mu$K, 100 $\mu$K, and 1 mK.
The higher temperature 100 $\mu$K spectrum is dominated by broad $d$-wave
components, on top of which a sharp subthermal $s$-wave component is clearly seen.
This is because there is a node in the $s$-wave scattering wave function at the Condon point
at relatively low scattering energy, unlike for other partial waves.
This sharp structure is a nice example of the subthermal line shapes discussed
by Machholm {\it et al}. \cite{machholm02} in the context of alkaline earth photoassociation.
Increasing the temperature to 1 mK leads to a quasi-continuum as several partial waves contribute, as seen
from Fig. \ref{fig6}.

The examples in this section show that the photoassociation spectra of various isotopes of the same species
can differ qualitatively. This variety is directly connected with the quantum defect
in the ground electronic state which is dependent on reduced mass and scattering length of the colliding
species.  This sensitivity of the rotational structure of photoassociation lines can help in the
determination or verification of the scattering length, as it was recently done in case of calcium
by Vogt {\it et al}. \cite{vogt07}.

\begin{figure}
\includegraphics[angle=0,width=\columnwidth,clip]{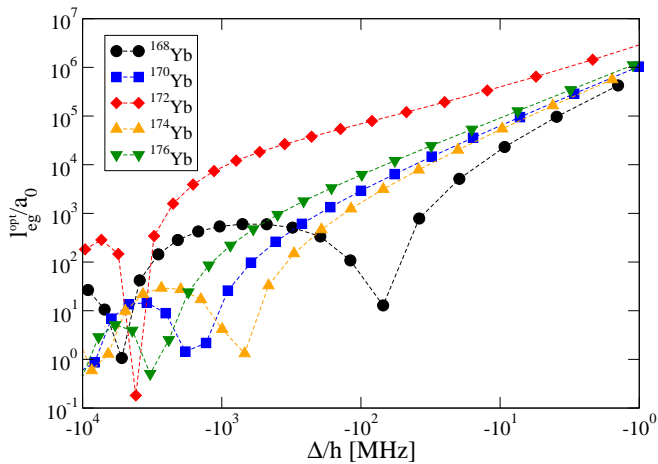}
\caption{(Color online)
The optical lengths $l^{\rm opt}_{eg}$ of the $s$-wave photoassociation features as a function of the
detuning $\Delta$ of the resonance from the atomic transition for the several bosonic isotopes of Yb.
The optical length determines the strength of an optical Feshbach resonance.  The optical length varies
linearly with laser intensity $I$ and these values have been normalized to $I=1\;{\rm W/cm^{2}}$
for $\varepsilon_{r}=0$.}
\label{fig7}
\end{figure}

A realistic description of near threshold interaction in the ground and excited electronic
states of the Yb$_{2}$ molecule is possible thanks to the experimental data \cite{tojo06,kitagawa08}
and the calculations we have shown here.
Our models of the interaction potentials can be used to predict the magnitude
of the optical lengths in Eq.~(\ref{equ6}) for the set of optical Feshbach resonances corresponding to the photoassociation
lines of the different Yb bosonic isotopes (fermionic isotopes are more complicated because of hyperfine
structure in the excited state~\cite{enomoto08}). Figure \ref{fig7} shows our calculated optical lengths
for the homonuclear bosonic isotopes of Yb.    The isotope $^{172}$Yb offers the strongest optical Feshbach
resonance because of its large negative scattering length, which enhances the amplitude of the $s$-wave
ground state wave function in the region of the Condon points.  Bound levels near the excited state
threshold have optical lengths at 1 W cm$^{-2}$ on the order of $10^6$ a$_0$, similar to a value measured
for a weakly bound excited level of $^{88}$Sr~\cite{zelevinsky06}.  This suggests that these resonances may
be of practical use for changing the scattering length for useful time scales while reducing spontaneous
emission loss processes by using large detuning.  Enomoto {\it et al.}~\cite{enomoto08ofr} have
experimentally demonstrated the possibility of some degree of optical Feshbach control in ultracold Yb gases.
Our predicted optical lengths should be helpful in choosing good resonances for controlling ultracold Yb
collisions by light.

\section{Conclusion}

One could naively  think  that this is a simple system so that not much difference should
be observed between the spectra of three bosonic isotopes like $^{172}$Yb, $^{174}$Yb, and $^{176}$Yb.
However, our calculations show that for temperatures about 100 $\mu$K the spectra for various isotopes
are qualitatively different. This difference is manifested by the fact that different rotational
lines are apparent in different isotopes. Moreover the shapes of photoassociation
lines are also affected. Some lines have subthermal widths depending on the isotope.
These differences can be well understood as due to the isotopic variation in the properties of the scattering
wave function in the ground electronic state. There is a clear connection between shape resonances in
ground scattering states and the appearance of strong lines coming from higher partial waves at relatively
low temperature of the order 100 $\mu$K. However, this only happens for excited bound states that have sufficiently short-range Condon points inside the location of the centrifugal barrier.
On the other hand, near-threshold bound states with long-range Condon points outside the centrifugal barrier do not show resonance enhancement.  For example, we showed
that the presence of $J_{e}=3$ lines for $^{172}$Yb at small detunings does not indicate that
a $d$-wave resonance occurs.  Thus, a good experimental method to determine whether resonances are present or not would be to measure photoassociation lines for more deeply bound excited states that have turning points near to or less that the ground state centrifugal barrier.  The existence of resonances correlates with the approximate value of the $s$-wave scattering length.

The intercombination photoassociation spectra of Yb are qualitatively different from spectra for
group II atoms such as Ca and Sr. For the case of Ca theory shows that the excited state potential
in the turning point range for levels near the dissociation limit is dominated by the van der Waals
potential, since the resonant dipolar interaction is so small \cite{ciurylo04,bussery05}.
The reflection approximation is not applicable in such a case \cite{ciurylo06}.
Strontium \cite{zelevinsky06} is a system which is half way between Ca and Yb.
Only the last three Sr bound states closest to the dissociation limit can be treated as dominated
by the resonant dipole interaction, while more deeply bound levels are determined mostly by
the van der Waals part of the interaction.
Photoassociation can be observed in Ca or Sr for two series of bound states
of $0_{u}^{+}$ and $1_{u}$ symmetry, respectively.  Although the resonant dipole interaction for the $1_{u}$
state is repulsive, its weakness allows it to be overcome by the attractive van der Waals interaction.
In contrast the Yb interaction in the excited electronic state with dissociation limit $^{1}S_{0}+^{3}P_{1}$
is dominated by the resonance interaction, so that only one potential curve of  $0_{u}^{+}$ symmetry
is attractive at long range and supports a series of detectable bound states.
The reflection approximation is well applicable in the case of $0_{u}^{+}$ levels of Yb.
Since the $1_{u}$ potential  becomes attractive at much shorter range of the interatomic
separation, the density of bound states near threshold is much less, and only one bound state with $1_{u}$
symmetry is expected near within a few GHz of threshold.

The heavier group IIb elements such as Cd and Hg will be similar to Yb.   For both species the interaction
in the excited state is dominated by the resonance interaction, and therefore one can expect that the
reflection approximation will be applicable.   Moreover both species have numerous stable isotopes.
Consequently, mass tuning of the scattering length should be applicable for both of these species,
and the results obtained in this work can be treated as universal and generic for Cd and Hg.

\begin{acknowledgments}
This work was partially supported by Grant-in-Aid for Scientific Research of JSPS
(18204035) and GCOE "The Next Generation of Physics,
Spun from Universality and Emergence" from MEXT of Japan.
PSJ was supported in part by the Office of Naval Research.
The research is part of the program of the National Laboratory FAMO in Toru\'n, Poland and
partially supported by the Polish MNISW (Project No. N N202 1489 33).
\end{acknowledgments}

\end{document}